\documentclass[prb,preprint,amsmath,amssymb]{revtex4}
\usepackage{color}
\begin{document}
\author{Kevin Leung$^*$}
\affiliation{Sandia National Laboratories, Albuquerque, New Mexico 87185,
USA; {\tt kleung@sandia.gov}}
\date{\today}
\title{Anomalous Ion Confinement Penalties and Giant
Ion-Screening Effects in One-Dimensional Nanopores}

\input epsf

\begin{abstract}

Nanoconfinement reduces the favorable hydration free energies of single 
ions, and is correlated with ion rejection and modified chemical reactivity
in water-filled nanopores.  Many factors contribute to the magnitude of
the observed confinement effect.  Here we use simple classical force fields
and non-polarizable carbon nanotubes filled with water as
minimal, ``hydrogen atom''-like models to evaluate the single-ion intrinsic
confinement hydration free energy penalty ($\Delta \Delta G_{\rm hyd}$).  In
tubes of radius $R$=7.5~\AA, we predict $\Delta \Delta G_{\rm hyd}$'s that
are up to 7.8~kcal/mol, are much larger for Cl$^-$ than the smaller
Na$^+$ ion, and contradict the canonical Born Equation for ion solvation.
Adding a 1.0~M background electrolyte reduces $\Delta \Delta G_{\rm hyd}$
for the Na$^+$/Cl$^-$ pair by an amount exceeding the Debye-H\"{u}ckel
estimate in unconfined media by almost an order of magnitude.  We identify
concentration-dependent ion-screening of confinement effects as a major,
unheralded consequence of electrolytes in cylindrical nanopores.

\end{abstract}

\maketitle

\setcounter{figure}{-1}
\begin{figure}
\centerline{\hbox{\epsfxsize=3.00in \epsfbox{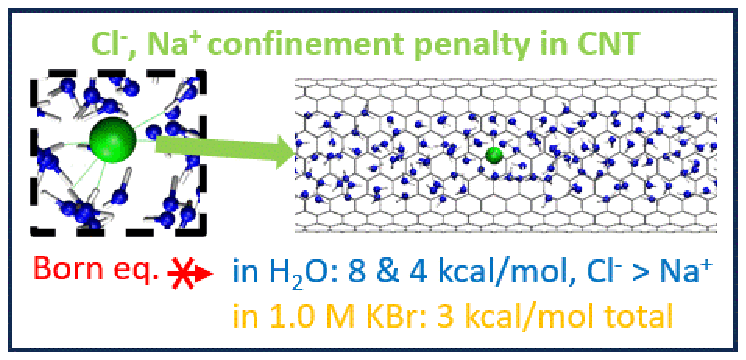} }}
\caption[]
{
TOC Graphic.
}
\end{figure}
\setcounter{figure}{0}

\clearpage
\newpage


It is widely accepted that the hydration free energies of single ions
($\Delta G_{\rm hyd})$ become less negative/favorable in confined liquid
media.\cite{review1,review2,ilgen0}  A confinement free energy penalty
($\Delta \Delta G_{\rm hyd}$, or $\Delta G_{\rm hyd}$ referenced to unconfined
systems) is qualitatively consistent with observed ion-rejection from porous
membranes,\cite{epsztein} increased ion-pairing,\cite{angel} and modified
surface adsorption\cite{mbonn} and chemical
reactivity,\cite{marx_review,ilgen0,ilgen2,ilgen3,ilgen4,paesani,paesani2,ysjun}
which are phenomena relevant to ion separation, extraction, desalination of
water, and energy
applications.\cite{mem_review,ionx,israel1,desal,bourg2,porous,salanne}

\begin{figure}
\centerline{\hbox{\epsfxsize=5.00in \epsfbox{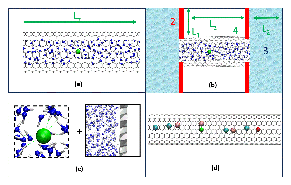} }}
\caption[]
{\label{fig1} \noindent
Different contributions to $\Delta \Delta G_{hyd}$ in CNT's.
(a) Intrinsic confinement effects, the focus of this work.
(b) Reservoir interface potential (red);  screening from reservoir (blue);
and finite size effects ($L_1$, $L_2$, $L_z$, green) which are eliminated by
the choice of our models.  The numbers correspond to the text.  (c) Reference
system with infinite tube curvature.  (d) Ion screening effect.  Blue,
white, gray, red, green, cyan, and pink represent O, H, C, Na, Cl, K, and Br
atoms, respectively.  Water is omitted from panel (d).
}
\end{figure}

A heuristic rationale for $\Delta \Delta G_{\rm hyd}$ is given by the
canonical, widely-cited Born Equation for ion solvation:\cite{born}
\begin{eqnarray}
\Delta G_{\rm hyd;B} &=& -(q^2/2a_{\rm eff}) (1-1/\epsilon); {\rm therefore} 
		\label{eq1} \\
\Delta \Delta G_{\rm hyd;B} &=& (q^2/2a_{\rm eff}) (1/\epsilon -1/\epsilon_w), 
		\label{eq2}
\end{eqnarray}
where $q$ is the ionic charge, $a_{\rm eff}$ is the effective ionic radius, 
$\epsilon_w$ and $\epsilon$ are the effective dielectric constants of the
unconfined reference and confined systems, respectively, and atomic units are
used.  If the relevant $\epsilon$ under confinement is less than $\epsilon_w$,
as has been widely
accepted,\cite{science,laage,gekle,netz2019,aluru1,aluru2,pham,ghoufi,bagchi,leung2023,bourg,leungho} 
$\Delta \Delta G_{\rm hyd}$$>$0, and the ion tends to be rejected from confined
spaces.  Since cations and anions are present in any electrolyte, elucidating
which ion is most unfavorable will significantly impact nanoscale applications.  
But many factors (Fig.~\ref{fig1}) are left out of Eqs.~\ref{eq1}-\ref{eq2}.
(1) The dielectric ($\epsilon$) response under confinement is
anisotropic,\cite{gekle,netz2019,aluru1,aluru2,leungho,pham,ghoufi,leung2023,bourg,bagchi}
and $\epsilon$-calculations via molecular dynamics (MD) simulations in
nano-cylinders are challenging even at the conceptual level.\cite{leungho} 
(2) Every interface in the system, e.g., that arising
from a water reservoir (Fig.~\ref{fig1}b),\cite{baldo2024,rempe,ho2025}
yields an additive contribution to $\Delta \Delta G_{\rm hyd}$.
This ``interface (or surface) potential'' contribution already
exists for unconfined water in contact with the vapor phase\cite{pratt89} 
and does not decay with distance from the interface.\cite{pratt89,surpot}
(3) Dielectric screening of the single ion by the reservoir (which exists
in many MD models\cite{baldo2024}) reduces $\Delta \Delta G_{\rm hyd}$.
Models used in such MD studies (e.g., Refs.~\onlinecite{baldo2024,rempe})
tend to have pore lengths which are too short, and water reservoirs too
close to the confined ion, compared to real-life, micron-long nanopores.
(4) The convergence of $\Delta \Delta G_{\rm hyd}$ with system sizes
has not been systematically explored (e.g., with respect to $L_1$, $L_2$, and
$L_z$ in Fig.~\ref{fig1}b).  These factors make MD-based interpretation of
confinement effects challenging.\cite{baldo2024,ilgen3}

In this work, we focus on one-dimensional pores.  Many naturally-occurring
and synthetic nanopores are cylindrical and microns in lengths, like
silica-based MCM-41, SBA-15 widely studied for nanoscience
applications,\cite{ilgen3} geochemically relevant imogolites,\cite{imogolite}
and carbon nanotubes (CNT).  The water-induced
``intrinsic confinement'' contribution to $\Delta \Delta G_{\rm hyd}$ in the
middle of these one-dimensional pores is unambiguous because the summed
Coulomb interaction converges absolutely in isolated nanotubes, providing a
confinement benchmark for Eqs.~\ref{eq1}-\ref{eq2}.  
Boundary contributions (Fig.~\ref{fig1}b) can be calculated and then added
separately.  Although molecular modeling of electrolytes under cylindrical
confinement has a long history,\cite{haymet} to our knowledge $\Delta
\Delta G_{\rm hyd}$'s for individual ions have yet to be elucidated 
\color{black} except in very narrow pores that disrupt the first hydration
shells.\cite{israel2} \color{black}

We will also show that adding a background 1.0~M electrolyte (henceforth
``ions'') strongly reduces $\Delta \Delta G_{\rm hyd}$ in narrow nanotubes.
This effect is not included in all modeling codes;\cite{oberhofer} some widely
used implicit solvent methods also use as input the pure solvent isotropic
$\epsilon$ but not the ionic strength.\cite{pcm}  $\sim$1.0~M ion
concentrations are relevant to sea water desalination,\cite{desal}
energy,\cite{porous,salanne} and nanopore applications (e.g., membrane-based
rejection of ions\cite{epsztein}) which exhibit regions with high local
ion concentrations.

We omit reservoirs and extrapolate the length of the single wall CNT to
infinity in order to pinpoint intrinsic confinement effects.  The 
\color{black} carbon-center-to-carbon-center \color{black} radii chosen
are $R$=7.5~\AA\, and 12.5~\AA, which are similar to those of imogolite
nanotubes.\cite{imogolite} We will show that these systems already generate
significant ion confinement effects; more extreme environments like
sub-nanometer diameter CNT's\cite{pham,israel2} or sub-nanometer
slits\cite{angel,paesani2} are not needed.  We apply thermodynamic integration
(TI) to the electrostatic contributions via MD simulations, 
\begin{equation}
\Delta G_{\rm hyd} = \int_0^1 d\lambda \langle d H(\lambda)/d\lambda)
		 \rangle_\lambda . \label{eq3}
\end{equation}
Here $\langle \rangle_\lambda$ represents thermal averaging in simulation
cells with a partially charged Na$^{\lambda+}$ or Cl$^{\lambda-}$ ion frozen
on the CNT axis.  This is the cylindrical confinement version of single
ion studies in unconfined liquid water.\cite{pratt96,pratt97,hunen}
Our CNT models are not polarizable or metallic, which would bestow different
dielectric behavior.\cite{korn2,korn3,netz2019,jimenez,sulpizi2}
The free energy of the same ion outside the tube is subtracted from
Eq.~\ref{eq3}.  This eliminates the monopole term\cite{makov} in
simulation cells with net charges.  In fact, $\Delta G_{\rm hyd}$ can be
calculated in different ways with different corrections,\cite{pratt96,sprik}
and they generally agree to $\sim$0.5~kcal/mol; see the Supporting Information
(SI) document which also describes the statistical uncertainties and
non-electrostatic contributions.  To obtain $\Delta
\Delta G_{\rm hyd}$ requires a reference state; here it is a single ion
removed from the liquid to infinity in vacuum, through a confining graphene
wall with vanishing curvature (Fig.~\ref{fig1}c) instead of the water-vapor
interface in other applications.\cite{pratt89}  This leads to a 8.9~kcal/mol
shift favoring cations and disfavoring anions for the force fields used.

\begin{figure}
\centerline{\hbox{\epsfxsize=5.00in \epsfbox{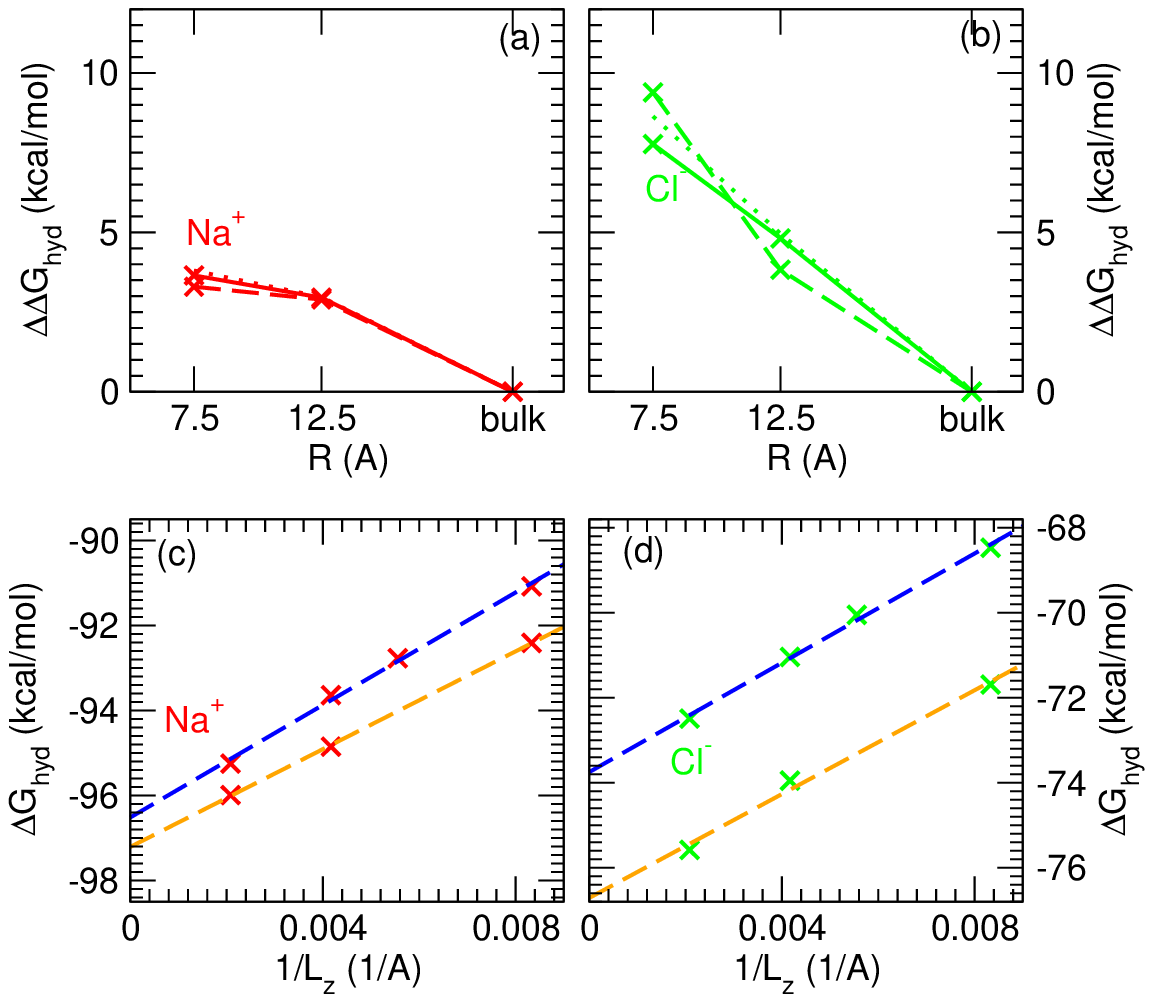} }}
\caption[]
{\label{fig2} \noindent
(a)-(b) $\Delta \Delta G_{\rm hyd}$ for Na$^+$ (red) and Cl$^-$ (green)
at different $R$, respectively, highlighting the anomalously large
$\Delta \Delta G_{\rm hyd}$ for Cl$^-$ at $R$=7.5~\AA.  
Dashed line: electrostatic contribution only; solid: total; dotted: 
total minus Lennard-Jones interaction with CNT. (c)-(d) Extrapolation
of $\Delta G_{\rm hyd}$ to infinite pore length for Na$^+$ and Cl$^-$,
respectively.  $R$=7.5~\AA\, and $R$=12.5~\AA\, are in blue and orange.  
}
\end{figure}

Figure~\ref{fig2}a-b demonstrates that, in a $R$=7.5~\AA\, radius CNT, $\Delta
\Delta G_{\rm hyd}$ values in confined water are substantial (3.7-7.8~kcal/mol)
compared to $k_{\rm B}T$. These effects, which should be readily detected in
experiments, are highly asymmetric with respective to cation/anion exchange.
$\Delta \Delta G_{\rm hyd}$ for Cl$^-$ decreases from 7.8~kcal/mol to
4.8~kcal/mol as $R$ increases from 7.5~\AA\, to 12.5~\AA.  In contrast, 
$\Delta \Delta G_{\rm hyd}$ for Na$^+$ only drops from 3.7 to 3.0~kcal/mol.  

The $\Delta G_{\rm hyd}$'s for confined ions used to calculate 
$\Delta \Delta G_{\rm hyd}$ have been extrapolated to 
$L_z$$\rightarrow$$\infty$ (Fig.~\ref{fig2}c-d).  The $\Delta G_{\rm hyd}$ 
differences between $L_z$=118~\AA\, and $L_z$$\rightarrow$$\infty$ are
5.4, 4.7, 5.3, and 5.2~kcal/mol for Na$^+$ at $R$=7.5~\AA\, and 
$R$=12.5~\AA, and Cl$^-$ in the two CNT's, respectively.
The scaling/convergence behavior in Fig.~\ref{fig2}c-d involves Ewald sums
and is not directly measurable.  In the SI, we apply real-space summation
of a lattice model which is more relevant to experimental $\Delta G_{\rm hyd}$
as a function of pore length. We also predict a quasi-(1/$L_z$) scaling regime
there, and estimate that a $L_z$=30~\AA\, tube model\cite{rempe} overestimates
the monovalent ion $\Delta \Delta G_{\rm hyd}$ by $\sim$3~kcal/mol compared
to $L_z$$\rightarrow$$\infty$.  This long-range requirement may impact the
fitting of machine learning force fields, and may require
explicit Coulomb forces be introduced.\cite{angel,cheng2025,ml_new}

\begin{figure}
\centerline{\hbox{\epsfxsize=5.00in \epsfbox{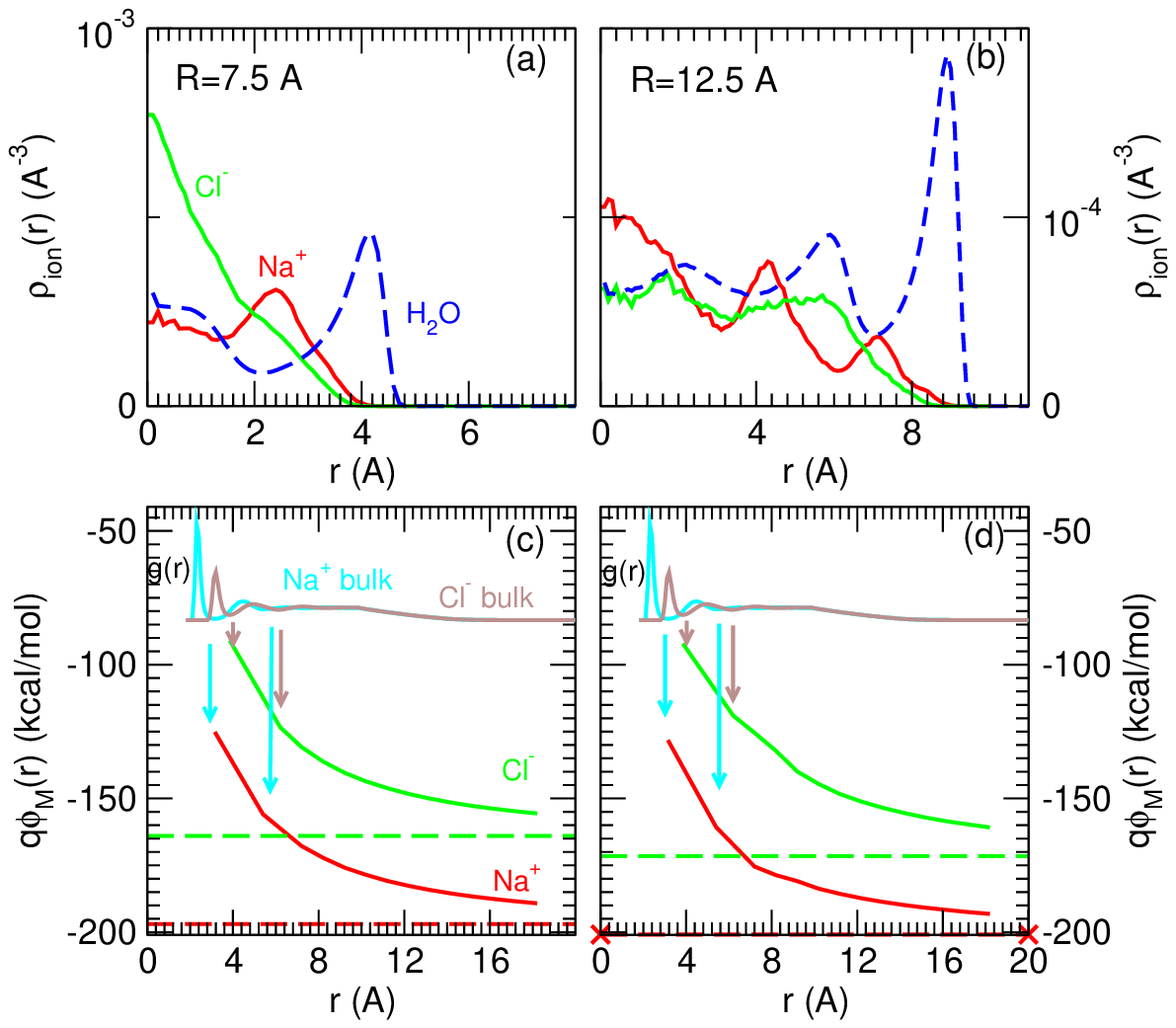} }}
\caption[]
{\label{fig3} \noindent
(a)-(b) Radial distribution of Na$^+$/Cl$^-$ for $R$=7.5~\AA\, and
$R$=12.5~\AA, respectively. Each cell contains either a Na$^+$ (red) or
a Cl$^-$ (green).  \color{black} $L_z$=118~\AA.  Water density profiles (blue)
are scaled by 1/200 and 1/500 in the two panels, respectively. \color{black}
(c)-(d) Madelung potential at $R$=7.5~\AA\, and $R$=12.5~\AA, respectively,
delimited by molecular cut-offs ($r$).  Red/blue denote values for confined 
Na$^+$/Cl$^-$.  Dashed lines are $\phi_M(r)$ with $r$$\rightarrow$$\infty$.
The $g(r)$'s of bulk water with the ions are also depicted as guides.
}
\end{figure}

Next we allow sampling of ion positions in the $x$-$y$ plane away from the
CNT axis.  The resulting radial ion distributions ($\rho_{\rm ion}(r)$)
depicted in Fig.~\ref{fig3}a-b show that Cl$^-$ is monotonically more
favored at the pore axis at $R$=7.5~\AA\, while Na$^+$ prefers regions where
the local water density is lowest.  Hence the large $\Delta \Delta G_{\rm hyd}$
for Cl$^-$ in the $R$=7.5~\AA\, CNT is not due to lowered Cl$^-$ probability
at the pore center compared to Na$^+$.  Converting $\rho_{\rm ion}(r)$ to
Boltzmann weights relative to $\rho_{\rm ion}$($r$=0) and integrating over
$r$ yield $\Delta G_{\rm hyd}$ configuration corrections of +0.51
and +1.27~kcal/mol for Na$^+$ and Cl$^-$ at R=7.5~\AA, and +0.84 and
+0.53~kcal/mol at R=12.5~\AA, respectively, after assuming tube volumes
are given by the number of H$_2$O present multipled by a molecular
volume of 29.9~\AA$^3$.  These corrections are a small fraction of the 
$\Delta \Delta G_{\rm hyd}$ reported above; they would have made the
Na$^+$/Cl$^-$ asymmetry in the $R$=7.5~\AA\, CNT even larger.  Here we
omit them because they would add non-intrinsic confinement effects (e.g.,
close-proximity ion interactions with CNT interior surfaces) to the
$\Delta \Delta G_{\rm hyd}$'s.  This point highlights the need to have
models with $\Delta \Delta G_{\rm hyd}$$>$1~kcal/mol for proof-of-principle
studies to overcome computational uncertainties.

\color{black} The $R$=7.5~\AA\, behavior (Fig.~\ref{fig3}a) is in best
agreement with the 1.15~nm diameter CNT results for Na$^+$ and Cl$^-$ 
in Ref.~\onlinecite{guo2022}. This is because, in our work, $R$ is measured
from the center of one CNT carbon atom to another without subtracting
the diameter of carbon atoms.  The force fields used are also different
from those in Ref.~\onlinecite{guo2022}.  \color{black}

$\Delta \Delta G_{\rm hyd}$ is predicted to be smaller for Na$^+$ than the
larger Cl$^-$.  This contradicts the canonical Born Equation.  The ionic
radii of Na$^+$ and Cl$^-$ are often quoted as 0.95~\AA\, and 1.81~\AA.
Even if one fits to the unconfined $\Delta G_{\rm hyd}$ without the boundary
term (Fig.~\ref{fig1}c), the Na$^+$ and Cl$^-$ $a_{\rm eff}$ would at best be 
similar (1.84 and 1.83~\AA).  According to Eq.~\ref{eq2}, if $\epsilon$=13.3
in the $R$=7.5~\AA\, CNT,\cite{leungho} $\Delta \Delta G_{\rm hyd:B}$ for
Na$^+$ and Cl$^-$ would be 5.6 and 5.7~kcal/mol respectively with these
$a_{\rm eff}$ instead of 3.7 and 7.8~kcal/mol.

Next we compute the restricted Madelung potential, $\phi_{\rm M}(r)$,
to examine the origin of the Na$^+$/Cl$^-$ asymmetry.  
\begin{equation}
\phi_{\rm M}(r)=\langle c_{\rm M} \sum_i' q_i/
                (\epsilon_w |{\bf r}_a-{\bf r}_i|)
                \rangle . \label{made}
\end{equation}
Here ${\bf r}_a$ is the position of the Na$^{\lambda+}$ or Cl$^{\lambda-}$,
the sum is over point charges $q_i$ of the classical force field model with
minimum image convention and molecular cut-off subject to the constraint that
the distance between the oxygen site of the H$_2$O being summed and the ion
is less than $r$, atomic units are used, and $c_{\rm M}$=$\pm$$\lambda$ for
Na$^{\lambda+}$/Cl$^{\lambda-}$ to make $\phi_{\rm M}(r)$ negative definite.

Figs.~\ref{fig3}c-d, depicted for $\lambda$=1, show that the magnitudes of
the $\phi_{\rm M}(r)$'s
for Cl$^-$ at $r$$<$3.85~\AA\, and $r$$<$6.20~\AA\ are in fact larger (more
favorable) for the $R$=7.5~\AA\, CNT than $R$=12.5~\AA.  Results for other
$\lambda$ values are qualitatively similar.  This shows, counter-intuitively,
that the anomalously large Cl$^-$ $\Delta \Delta G_{\rm hyd}$ at $R$=7.5~\AA\,
does not arise from short-range interactions.  By elimination, it
must be due to competition among long-range effects: (1) dielectric solvation
(agnostic to the sign of the charge); (2) the interactions with preferentially
oriented water at the CNT interior surface causing the 8.9~kcal/mol
interfacial potential favoring cations; and (3) CNT curvature which modifies
this interfacial potential.  Asymmetry between cations and anions has also
been noted for contact ion pairs in slit pores\cite{fenter} and other
contexts.\cite{schenter}

Finally, we add a 1.0~M KBr background electrolyte in the $R$=7.5~\AA\, CNT.
Wider CNT's exhibit smaller confinement effects in 1.0~KBr,\cite{leungho}
and are not considered.  Once again we use the Born formula (Eq.~\ref{eq1})
as starting point.  \color{black} 
It has often been reported that the water dielectric
constant $\epsilon_w$ is reduced in the presence of ions.\cite{klein_diel} 
If adding ions reduces $\epsilon_w$ and $\epsilon$ under confinement
proportionally, $\Delta \Delta G_{\rm hyd}$ should increase. However, such
decreased $\epsilon_{\rm w}$ and $\epsilon$ are calculated/measured at MHz/GHz
frequencies, while ion hydration corresponds to zero frequency at which
$\epsilon$ of the medium diverges for electrolytes.\cite{leung2023}  In
our recent MD simulations, a 0.5~M background electrolyte appears to
eliminate the $\Delta \Delta G_{\rm hyd}$ predicted in water confined in a
nanoslit.\cite{baldo2024}  We will show that adding 1.0~M KBr also decreases
$\Delta \Delta G_{\rm hyd}$ here, qualitatively consistent with setting
$\epsilon$=$\infty$ in Eq.~\ref{eq1}. \color{black}

\begin{figure}
\centerline{\hbox{\epsfxsize=5.00in \epsfbox{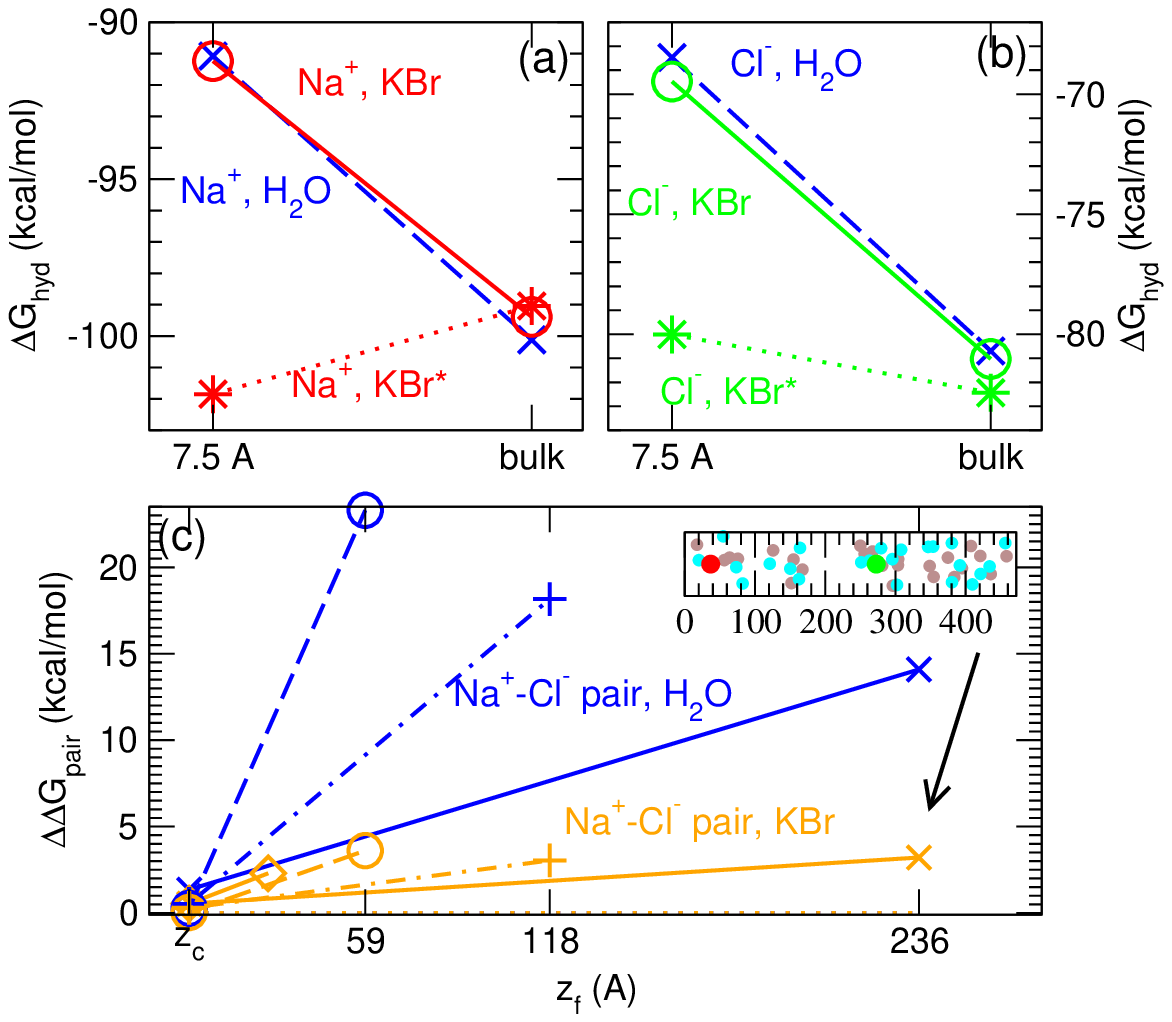} }}
\caption[]
{\label{fig4} \noindent
(a)-(b): $\Delta G_{\rm hyd}$ for isolated Na$^+$ and Cl$^-$, respectively,
in the $R$=7.5~\AA\, CNT.
Blue/crosses: in water. Solid red or green with circles: in 1.0~M KBr with
equal number of cations and anions. Dotted red or green with stars: in
1.0~M KBr with one less cation or anion.  $L_z$=118~\AA.  These panels show
that single-ion calculations are ambiguous at finite ion concentrations.
(c) $\Delta \Delta
G_{\rm pair}(z)$ for Na$^+$+Cl$^-$ ion pair in contact ($z$=$z_c$)
or half a box length away ($z$=$z_f$=$L_z$/2) in 1.0~M KBr (orange), and sum
of $\Delta G_{\rm hyd}$ for isolated Na$^+$ plus Cl$^-$ in water (blue, from
Fig.~\ref{fig2}c-d).  Solid/cross, dot-dashed/plus, and dashed/circle are for
$L_z$=472~\AA, $L_z$=236~\AA, and $L_z$=118~\AA.  The diamond
is for $L_z$=472~\AA, $z$=28~\AA.  Inset: ion segregation in one MD snapshot.
Red, green, brown, and cyan are Na$^+$, Cl$^-$, K$^+$, and Br$^-$.
}
\end{figure}

Our approach used for pure water needs to be modified for electrolytes.  To
illustrate this, consider the \color{black} single-ion TI $\Delta G_{\rm hyd}$
\color{black} in Fig.~\ref{fig4}a-b which correspond to 1.0~M KBr 
electrolytes in the $L_z$=118~\AA\, cell.  If the numbers of K$^+$ and Br$^-$
are equal, and the cell is charge-neutral at $\lambda$=0 during TI (but not
at any other $\lambda$, Eq.~\ref{eq3}), the single-ion $\Delta G_{\rm hyd}$'s
are almost the same at 0.0~M or 1.0~M KBr for the \color{black} confined Na$^+$ 
(Fig.~\ref{fig3}a) and the confined Cl$^-$ (Fig.~\ref{fig3}b). \color{black}
If the cell has one less K$^+$ or Br$^-$ and is charge-neutral at $\lambda$=1,
the \color{black} sum of single-ion \color{black}
$\Delta G_{\rm hyd}$ becomes $\sim$21.2~kcal/mol more favorable.  In contrast,
in unconfined electrolytes, whether the numbers of K$^+$ and Br$^-$ are equal
or not only has a $\sim$1~kcal/mol effect on $\Delta G_{\rm hyd}$.  
\color{black} The reason is that the canonical ensemble we use suppresses charge
fluctuations,\cite{widom2} \color{black}
which is particularly significant if the net charge is concentrated inside
the CNT, not smeared across the entire cell. In principle, Grand Canonical
Monte Carlo (GCMC) can be used to remove this ambiguity.  GCMC can add
counter-ions and make the simulation cell probabilistically charge-neutral,
but this adds a significant computational cost.\cite{luzar}

To circumvent this ambiguity, we place a Na$^{\lambda+}$ and a Cl$^{\lambda-}$
ion on the CNT axis either at contact ($z_c$=2.75~\AA\, apart) or $z_f=L_z/2$
apart from each other at 1.0~M KBr concentration, and simultaneously charge
both ions to $\lambda$=$\pm$1 using TI.  Now the system is charge-neutral
over the entire $\lambda$-path.  

The resulting \color{black} contact-ion-pair \color{black}
$\Delta \Delta G_{\rm pair}(z_c)$ are first calculated in
confined water and confined 1.0~M KBr (Fig.~\ref{fig4}c).  They are within
0.3-1.4~kcal/mol of the unconfined $\Delta \Delta G_{\rm hyd}$ for Na$^+$ and
Cl$^-$ combined.  To a first approximation, these uncharged contact ion pairs
(CIP's) do not incur monopole confinement penalties (Eqs.~\ref{eq1}-\ref{eq2}).

In 1.0~M KBr, the confinement penalties for well-separated Na$^+$/Cl$^-$
are drastically reduced from those in water, taken as sum of Na$^+$ and Cl$^-$
contributions from Fig.~\ref{fig2}c-d (Fig.~\ref{fig4}c).  They are also less
$L_z$-dependent. When the Na$^+$ and Cl$^-$ are separated by $z$=28~\AA\,
at $L_z$=472~\AA\, (Fig.~\ref{fig4}c), $\Delta \Delta G_{\rm pair}(z)$ is
already comparable to that at $z$=$L_z$/2.  Even though our ion pair approach
cannot separate the Na$^+$ and Cl$^-$ contributions, the results strongly
suggest that $\Delta \Delta G_{\rm hyd}$ for both ions are much reduced.  
\color{black} We have further repeated the 
calculations in the $L_z$=472~\AA\, simulation cell at a lower, 0.5~M KBr
concentration; $\Delta \Delta G_{\rm pair}(z_f)$ is very similar (lower by
0.21~kcal/mol) compared with the 1.0~M KBr simulation.
\color{black}

In other words, the KBr ions {\it screen} most of the water-Na$^+$/Cl$^-$
interactions that lead to the confinement penalty.  This behavior can be
anticipated from and reconciled with previous predictions of background ions
drastically reducing the free energy cost of pulling apart contact ion pairs
inside this CNT.\cite{leungho} In solid state materials, all mobile charges
are recognized to screen all other charges.  In unconfined liquids,
the role ions play in screening is sometimes omitted.\cite{pcm,oberhofer}  At
low ion concentrations, the ion screening contribution can be estimated using
the Debye-H\"{u}ckel \color{black} (DH) \color{black}
approach.\cite{npj,boda2023,henriksen,netz2020}
The correction to $\Delta G_{\rm hyd}$ is \color{black}
$-q^2/[2\epsilon r_D ( 1 + a_{\rm eff}/r_D)]$ \color{black} in
atomic units in unconfined electrolytes, where $r_D$ is the Debye screening
length\cite{npj,boda2023,henriksen} \color{black} and $q$=$\pm$1 is the ionic
charge.  As $r_D$$\propto$$\epsilon^{0.5}$, this attractive ion-ion interaction
$\propto$$1/\epsilon^{1.5}$; other formulations exhibit more complex,
cross-over scaling behavior.\cite{boda2021} \color{black}
Using $a_{\rm eff}$=1.84~\AA\, (which is force field-dependent) and taking
$\epsilon$=60 for an unconfined 1.0~M monovalent electrolyte,\cite{klein_diel}
the DH approach
yields a correction of -1.2~kcal/mol for Na$^+$ + Cl$^-$.  In contrast, in our
$L_z$=472~\AA, $R$=7.5~\AA\, cell, the 1.0~KBr reduces $\Delta \Delta
G_{\rm pair}$ by \color{black} 8.2~to~9.1~kcal/mol depending on whether
the Na$^+$ and Cl$^-$ are placed 236 or 28~\AA\, apart \color{black} -- almost
an order of magnitude larger.  This suggests that more quantitative solvation
theories\cite{oberhofer,mundy1} than \color{black} DH \color{black} need to be
developed to quantify confinement in electrolytes, particularly because both
$\epsilon$ and $r_D$ may be anisotropic in nanopores.  \color{black} This
finding may also be relevant to unconfined battery electrolytes with low
solvent $\epsilon$'s.\cite{borodin,gallant} \color{black}

In the largest ($L_z$=472~\AA) cell, KBr aggregation behavior is observed; 
gaps without ions appear.  Furthermore, the 1.0~M KBr drastically decreases
the favorable ion-pairing free energy among the K$^+$ and Br$^-$ themselves,
and decreases the persistence times of K-Br CIP's.  See the SI for details.
Nevertheless, more K$^+$/Br$^-$ CIP's (50\%) are formed inside the 
$R$=7.5~\AA\, CNT than in the unconfined electrolyte (33\%), likely due to
the residual confinement effect in 1.0~KBr (Fig.~\ref{fig4}c).  
For a quantitative comparison with ion-pairing experiments,\cite{ilgen3}
lower electrolyte concentrations, multivalent ions, the dielectric materials
outside the CNT, charges on nanopore surfaces, more accurate force fields,
and CNT polarizability need to be considered.  We defer those to future work.

In conclusion, we have calculated the intrinsic confinement free energy
penalties ($\Delta \Delta G_{\rm hyd}$) of single ions in water-filled
non-polarizable CNT's.  In a $R$=7.5~\AA\, CNT, $\Delta \Delta
G_{\rm hyd}$$\sim$3.7 and~7.8~kcal/mol for Na$^+$ and Cl$^-$, respectively.
Na$^+$ is strongly favored over Cl$^-$ in this CNT, contrary to
what the canonical Born Equation\cite{born} would predict, suggesting the
Born formulation needs to be modified for strongly confined media.  
$\Delta \Delta G_{\rm hyd}$ converges slowly with CNT length ($L_z$), depending
on the computational method used; at $L_z$=472~\AA, it is within 1~kcal/mol of
the value extrapolated to infinity.  By considering a Na$^+$/Cl$^-$ pair
far apart along the CNT axis, we show that adding 1.0~M KBr significantly
reduces $\Delta \Delta G_{\rm hyd}$.  This emphasizes that ions strongly
screen water-ion interactions, of which $\Delta \Delta G_{\rm hyd}$ in
confined water is one manifestation.  Inferring ion solvation properties from
the dielectric properties of pure, confined water alone\cite{science,laage,gekle,netz2019,aluru1,aluru2,pham,ghoufi,bagchi,leung2023,bourg,leungho} 
would omit this crucial contribution.  We identify finite-concentration
ion-screening as a major, unheralded consequence of confined electrolytes
in cylindrical nanopores.  

\section*{Supporting Information Available}

The Supporting Information is available free of charge at
https://pubs.acs.org/.
Discussion of computational details; convergence tests; supplemental results; 
lattice model real space summation.

\section*{Acknowledgement}

We thank Dusan Bratko, Marialore Sulpizi, Jeff Greathouse, Ward Thompson, 
Anastasia Ilgen, and Tuan Ho for valuable suggestions.
This work is based on materials support by the U.S. DOE Office of Basic
Energy Sciences, Division of Chemical Sciences, Geosciences, and Biosciences
under Field Work Proposal Number 24-015452 at Sandia National Laboratories.
This article has been authored by an employee of National Technology \&
Engineering Solutions of Sandia, LLC under Contract No.~DE-NA0003525 with
the U.S.~Department of Energy (DOE). The employee owns all right,
title and interest in and to the article and is solely responsible for
its contents. The United States Government retains and the publisher,
by accepting the article for publication, acknowledges that the United
States Government retains a non-exclusive, paid-up, irrevocable, world-wide
license to publish or reproduce the published form of this article or
allow others to do so, for United States Government purposes. The DOE
will provide public access to these results of federally sponsored research
in accordance with the DOE Public Access Plan
https://www.energy.gov/downloads/doe-public-access-plan. This paper
describes objective technical results and analysis. Any subjective views
or opinions that might be expressed in the paper do not necessarily represent
the views of the U.S.~Department of Energy or the United States Government.

\newpage

\noindent{\bf {\large References}}

\newpage

\end{document}